\documentclass[%draft,published,
notoc,nohyper]{JHEP} % 10pt is ignored!

\usepackage[psamsfonts]{amsfonts}

\newcommand{\tr}{\mathop{\rm tr}\nolimits}

\newcommand{\U}{\mathop{\rm {}U}}
\newcommand{\rmd}{{\rm d}}
\newcommand{\sgn}{\mathop{\rm sgn}}
\newcommand{\vol}{\mathop{\rm vol}}
\newcommand\fverb{\setbox\pippobox=\hbox\bgroup\verb}
\newcommand\fverbdo{\egroup\medskip\noindent%
                        \fbox{\unhbox\pippobox}\ }
\newcommand\fverbit{\egroup\item[\fbox{\unhbox\pippobox}]}
\newbox\pippobox
\title{
Axial anomaly with the overlap-Dirac operator in arbitrary dimensions}

\author{Takanori Fujiwara\\
Department of Mathematical Sciences, Ibaraki University, Mito 310-8512, Japan\\
E-mail: \email{fujiwara@mx.ibaraki.ac.jp}}
\author{Keiichi Nagao\\
High Energy Accelerator Research Organization (KEK), Tsukuba 305-0801, Japan\\
E-mail: \email{nagao@post.kek.jp}}
\author{Hiroshi Suzuki\\
Department of Mathematical Sciences, Ibaraki University, Mito 310-8512, Japan\\
E-mail: \email{hsuzuki@mx.ibaraki.ac.jp}}
\received{\today}               %%
\accepted{\today}               %% These are for published papers.
\preprint{IU-MSTP/52\\KEK-TH-838\\\heplat{0208057}}
\abstract{
We evaluate for arbitrary even dimensions the classical continuum limit of the
lattice axial anomaly defined by the overlap-Dirac operator. Our calculational
scheme is simple and systematic. In particular, a powerful topological argument
is utilized to determine the value of a lattice integral involved in the
calculation. When the Dirac operator is free of species doubling, the
classical continuum limit of the axial anomaly in various dimensions is
combined into a form of the Chern character, as expected.
}

\begin{document} 

\maketitle %%%%%%%%%% THIS IS IGNORED %%%%%%%%%%%

In this note, we evaluate  for arbitrary even dimensions~$d=2n$ the classical
continuum limit of the lattice axial anomaly defined by the overlap-Dirac
operator~\cite{Neuberger:1998fp}. This quantity has been studied many
times~\cite{Kikukawa:1998pd}--\cite{Frewer:2000ee}; but all treat $d=4$~case.
There also exist general
arguments~\cite{Fujikawa:1998if,Luscher:2000un,Reisz:1999cm,Frewer:2000ee} that
the classical continuum limit reproduces the axial anomaly in the continuum
theory, when the Dirac operator is free of species doubling. If it is possible,
however, it is certainly preferable to demonstrate this by an explicit
calculation. We do this for arbitrary dimensions and obtain the expected
result; the classical continuum limit of the axial anomaly in various
dimensions is combined into a form of the Chern
character~\cite{Alvarez-Gaume:1985ex} [see~eq.~(\ref{fortyone})]. Our
calculation consists of two steps: First we determine $O(a^0)$~terms in the
axial anomaly without expanding it with respect to the gauge coupling; this
scheme is similar to that of~refs.~\cite{Seiler:1981jf}--\cite{Haga:1996at},
\cite{Adams:1998eg,Suzuki:1998yz,Fujikawa:2000qw}. Next, we evaluate a lattice
integral involved; here a powerful topological argument is utilized to
determine its value (closely related topological arguments can be found
in~refs.~\cite{Coste:1989wf,Golterman:1992ub,Kikukawa:1998pd}). A combination
of these two schemes provides a quite simple and systematic method to evaluate
the classical continuum limit of the axial anomaly. To illustrate this point,
we present a calculation of the axial anomaly with the Wilson-Dirac operator
for arbitrary dimensions in appendix~A.

The overlap-Dirac operator~\cite{Neuberger:1998fp} is defined
by\footnote{Our notations: $a$ denotes the lattice spacing. Greek letters,
$\mu$, $\nu$, \dots\ run from $1$ to~$d=2n$. Repeated indices are understood to
be summed over, unless noted otherwise.
$\{\gamma_\mu,\gamma_\nu\}=2\delta_{\mu\nu}$,
$\gamma_\mu^\dagger=\gamma_\mu$
and~$\gamma_{d+1}=(-i)^n\gamma_1\cdots\gamma_d$;
$\gamma_{d+1}^2=1$ and~$\gamma_{d+1}^\dagger=\gamma_{d+1}$ follow from this.}
\begin{eqnarray}
   &&D={1\over a}\left[1-A(A^\dag A)^{-1/2}\right],\qquad
   A=m_0-aD_{\rm w},
\label{one}\\
   &&D_{\rm w}={1\over2}\left[\gamma_\mu(\nabla_\mu^*+\nabla_\mu)
   -ar\nabla_\mu^*\nabla_\mu\right],
\label{two}
\end{eqnarray}
where $m_0$ and~$r$ are free parameters. We assume $r>0$ in what follows. In
the absence of the gauge field, the Dirac operator is free of species doubling
if $0<m_0/r<2$. $\nabla_\mu$ and $\nabla_\mu^*$ in this expression are forward
and backward covariant difference operators respectively\footnote{The link
variable~$U(x,\mu)$ denotes a matrix in a (in general reducible) unitary
representation of the gauge group, to which the fermion belongs.}
\begin{eqnarray}
   \nabla_\mu\psi(x)&=&{1\over a}
   \left[U(x,\mu)\psi(x+a\hat\mu)-\psi(x)\right],
\label{three}\\
   \nabla_\mu^*\psi(x)&=&{1\over a}
   \left[\psi(x)-U(x-a\hat\mu,\mu)^\dag\psi(x-a\hat\mu)\right].
\label{four}
\end{eqnarray}
By construction, the overlap-Dirac operator satisfies the Ginsparg-Wilson
relation $\gamma_{d+1}D+D\gamma_{d+1}=aD\gamma_{d+1}D$~\cite{Ginsparg:1982bj}.
Due to this relation, the lattice fermion
action~$S_{\rm F}=a^d\sum_x\overline\psi(x)D\psi(x)$ is invariant under the
modified chiral transformation~\cite{Luscher:1998pq}
\begin{equation}
   \delta\psi(x)=\epsilon\gamma_{d+1}\left(1-{1\over2}aD\right)\psi(x),\qquad
   \delta\overline\psi(x)=\overline\psi\left(1-{1\over2}aD\right)
   \gamma_{d+1}\epsilon.
\label{five}
\end{equation}
The fermion integration measure, however, acquires a non-trivial jacobian under
the transformation~\cite{Luscher:1998pq}
\begin{equation}
   \delta\prod_x\rmd\psi(x)\rmd\overline\psi(x)
   =-2\left[a^d\sum_x\epsilon q(x)\right]
   \prod_x\rmd\psi(x)\rmd\overline\psi(x),
\label{six}
\end{equation}
where
\begin{equation}
   q(x)={1\over a^d}\tr\left[
   \gamma_{d+1}\left(1-{1\over2}aD(x,x)\right)\right].
\label{seven}
\end{equation}
This framework thus realizes a desired breaking pattern of the chiral
Ward-Takahashi identity even with finite lattice spacings;\footnote{Note that
the jacobian is unity for flavor non-singlet chiral transformations for which
$\tr\epsilon=0$.} it is especially suitable for a study of phenomena associated
to the axial anomaly, such as the $\U(1)$
problem~\cite{Chandrasekharan:1998wg}--\cite{Giusti:2001xh}. The lattice axial
anomaly~$q(x)$ may also be regarded as the index density, because
$a^d\sum_x q(x)$ is an integer depending on the gauge-field
configuration~\cite{Hasenfratz:1998ri}.

We first note that a diagonal element of the kernel of an operator on the
lattice can be expressed as
\begin{eqnarray}
   {\cal O}(x,x)&=&\sum_y{\cal O}(x,y)\delta_{y,x}
   =\int_{-\pi}^\pi{\rmd^dk\over(2\pi)^d}\,e^{-ikx/a}
   \sum_y{\cal O}(x,y)e^{iky/a}
\nonumber\\
   &=&\int_{-\pi}^\pi{\rmd^dk\over(2\pi)^d}\,e^{-ikx/a}{\cal O}e^{ikx/a}.
\label{eight}
\end{eqnarray}
So we note
\begin{equation}
   \left\{{A\atop A^\dagger}\right\}e^{ikx/a}f(x)
   =e^{ikx/a}\left\{\mp\gamma_\mu(is_\mu+aQ_\mu)+m_0
   +r\left[\sum_\mu(c_\mu-1)+aR\right]\right\}f(x),
\label{nine}
\end{equation}
where
\begin{equation}
   s_\mu=\sin k_\mu,\qquad c_\mu=\cos k_\mu,
\label{ten}
\end{equation}
and
\begin{equation}
   Q_\mu={1\over2}(e^{ik_\mu}\nabla_\mu+e^{-ik_\mu}\nabla_\mu^*),\qquad
   R={1\over2}\sum_\mu
   (e^{ik_\mu}\nabla_\mu-e^{-ik_\mu}\nabla_\mu^*).
\label{eleven}
\end{equation}
{}From eqs.~(\ref{one}) and~(\ref{seven}), we thus have
\begin{eqnarray}
   q(x)&=&{1\over2a^d}\int_{-\pi}^\pi{\rmd^d k\over(2\pi)^d}
   \tr\Biggl[\gamma_{d+1}
   \left\{-\gamma_\mu(is_\mu+aQ_\mu)+m_0
   +r\left[\sum_\mu(c_\mu-1)+aR\right]\right\}
\nonumber\\
   &&\qquad
   \times\Biggl(-(is_\nu+aQ_\nu)^2
   +\left\{m_0+r\left[\sum_\nu(c_\nu-1)+aR\right]\right\}^2
\nonumber\\
   &&\qquad\qquad\qquad\qquad\qquad\quad
   -{a^2\over4}[\gamma_\nu,\gamma_\rho][Q_\nu,Q_\rho]
   +a^2r[\gamma_\nu Q_\nu,R]\Biggr)^{-1/2}\Biggr]1,
\label{twelve}
\end{eqnarray}
where we have used $\tr\gamma_{d+1}=0$ and
$\gamma_\nu\gamma_\rho=\delta_{\nu\rho}+[\gamma_\nu,\gamma_\rho]/2$. In this
form, an expansion of~$q(x)$ with respect to the lattice spacing~$a$ is
straightforward. In the classical continuum limit, the gauge
potential~$A_\mu(x)$ introduced by
\begin{equation}
   U(x,\mu)={\cal P}\exp\left[
   a\int_0^1\rmd t\,A_\mu(x+(1-t)a\hat\mu)\right],
\label{thirteen}
\end{equation}
is regarded as a smooth function of~$x$ of the order~$O(a^0)$. The operators
$Q_\mu$ and~$R$ are of~$O(a^0)$ in this limit:
\begin{eqnarray}
   &&Q_\mu=c_\mu(\partial_\mu+A_\mu)+O(a),\qquad\hbox{no sum over $\mu$},
\label{fourteen}\\
   &&R=\sum_\mu is_\mu(\partial_\mu+A_\mu)+O(a),
\label{fifteen}
\end{eqnarray}
and
\begin{eqnarray}
   &&[Q_\mu,Q_\nu]=c_\mu c_\nu F_{\mu\nu}+O(a),\qquad
   \hbox{no sum over $\mu$ and~$\nu$},
\label{sixteen}\\
   &&[Q_\mu,R]=ic_\mu\sum_\nu s_\nu F_{\mu\nu}+O(a),\qquad
   \hbox{no sum over $\mu$},
\label{seventeen}
\end{eqnarray}
where $F_{\mu\nu}=\partial_\mu A_\nu-\partial_\nu A_\mu+[A_\mu,A_\nu]$. Thus
it is straightforward to find $O(a^0)$~terms in eq.~(\ref{twelve}), because
the trace over Dirac indices requires at least $d$~gamma matrices:
\begin{equation}
   \tr\gamma_{d+1}\gamma_{\mu_1}\gamma_{\nu_1}\cdots
   \gamma_{\mu_n}\gamma_{\nu_n}
   =i^n2^n\epsilon_{\mu_1\nu_1\cdots\mu_n\nu_n}.
\label{eighteen}
\end{equation}
Namely, we have
\begin{eqnarray}
   q(x)&=&{1\over2}\int_{-\pi}^\pi{\rmd^d k\over(2\pi)^d}{-1/2\choose n}
   \left\{\sum_\rho s_\rho^2+\left[m_0+r\sum_\rho(c_\rho-1)\right]^2
   \right\}^{-n-1/2}
\nonumber\\
   &&\qquad\times\Biggl(
   \tr\left\{\gamma_{d+1}\left[m_0+r\sum_\sigma(c_\sigma-1)\right]
   \left(-{1\over2}\gamma_\mu\gamma_\nu c_\mu c_\nu F_{\mu\nu}\right)^n\right\}
\nonumber\\
   &&\qquad\qquad+\tr\Biggl\{\gamma_{d+1}\gamma_\sigma s_\sigma
   \Biggl[
   \left(-{1\over2}\gamma_\mu\gamma_\nu c_\mu c_\nu
   F_{\mu\nu}\right)^{n-1}
   r\gamma_\lambda c_\lambda s_\tau F_{\lambda\tau}
\nonumber\\
   &&\qquad\qquad\qquad\qquad\qquad\qquad\qquad\qquad
   +{\rm permutations}\,\Biggr]
   \Biggr\}\Biggr)+O(a).
\label{nineteen}
\end{eqnarray}
A little calculation using the identity~(\ref{eighteen}) shows
\begin{equation}
   q(x)={(-i)^n\over2(2\pi)^d}{-1/2\choose n}I(m_0,r)\,
   \epsilon_{\mu_1\nu_1\cdots\mu_n\nu_n}
   \tr\left(F_{\mu_1\nu_1}\cdots F_{\mu_n\nu_n}\right)+O(a),
\label{twenty}
\end{equation}
where the lattice integral~$I(m_0,r)$ is given by
\begin{equation}
   I(m_0,r)=\int_{\cal B}\rmd^dk\,{\cal I}(k;m_0,r),
\label{twentyone}
\end{equation}
with
\begin{eqnarray}
   {\cal I}(k;m_0,r)&=&\left(\prod_\mu c_\mu\right)
   \left\{\sum_\nu s_\nu^2+\left[m_0+r\sum_\nu(c_\nu-1)\right]^2
   \right\}^{-n-1/2}
\nonumber\\
   &&\qquad\qquad\qquad\qquad\times
   \left[m_0+r\sum_\rho(c_\rho-1)
   +r\sum_\rho{s_\rho^2\over c_\rho}\,\right],
\label{twentytwo}
\end{eqnarray}
and
\begin{equation}
   {\cal B}=\left\{k_\mu\in{\mathbb R}^d\bigm|
   -{\pi\over2}\leq k_\mu\leq{3\pi\over2}\right\}.
\label{twentythree}
\end{equation}
Here we have shifted the integration region
as~$k_\mu\in[-\pi,\pi]\to k_\mu\in[-\pi/2,3\pi/2]$ for later convenience.

We now turn to a calculation of the lattice integral~$I(m_0,r)$. We utilize the
following topological argument. We first introduce a mapping from the
Brillouin zone~${\cal B}$ to the unit sphere~$S^d$. The mapping is defined by
\begin{eqnarray}
   &&\theta_0=m_0+r\sum_\mu(c_\mu -1),
\label{twentyfour}\\
   &&\theta_\mu=s_\mu,\qquad{\rm for}\qquad\mu=1,\ldots,d,
\label{twentyfive}
\end{eqnarray}
and
\begin{equation}
   x_A={\theta_A\over\epsilon},\qquad\epsilon=\sqrt{\sum_A\theta_A^2},
\label{twentysix}
\end{equation}
where $x_A$ ($A=0$, $1$, $\ldots$, $d$) is the coordinate
of~${\mathbb R}^{d+1}$ in which the unit sphere~$\sum_Ax_A^2=1$ is embedded.
Since this mapping is periodic on the Brillouin zone~${\cal B}$, we can regard
${\cal B}$ as a torus~$T^d$. Namely $k_\mu\to x_A$ defines a
mapping~$f:T^d\to S^d$. The crucial observation is that the volume form on this
sphere coincides with the integrand of~$I(m_0,r)$:
\begin{eqnarray}
   \Omega&=&{1\over d!}\,
   \epsilon_{A_0\cdots A_d}x_{A_0}\rmd x_{A_1}\wedge\cdots\wedge\rmd x_{A_d}
\nonumber\\
   &=&{1\over d!}{1\over\epsilon^{d+1}}\,\epsilon_{A_0\cdots A_d}
   \theta_{A_0}\rmd\theta_{A_1}\wedge\cdots\wedge\rmd\theta_{A_d}
\nonumber\\
   &=&{\cal I}(k;m_0,r)\,\rmd k_1\wedge\cdots\wedge\rmd k_d.
\label{twentyseven}
\end{eqnarray}
This shows that the integral of~$\Omega$ on a (sufficiently small) coordinate
patch~$U$ on~$S^d$ is given by
\begin{equation}
   \int_U\Omega=\sgn\left[{\cal I}(k^j;m_0,r)\right]
   \int_{U^j}{\cal I}(k;m_0,r)\,\rmd k_1\wedge\cdots\wedge\rmd k_d,
\label{twentyeight}
\end{equation}
where $U^j$ ($j=1$, \dots, $m$) is a component of the inverse image of~$U$
under~$f$, $f^{-1}(U)$:
\begin{equation}
   f^{-1}(U)=U^1\cup\cdots\cup U^m\subset T^d,
\label{twentynine}
\end{equation}
and $k^j$ ($j=1$, \dots, $m$) is a certain point on~$U^j$. For a sufficiently
small~$U$, $U_j$ are pairwise disjoint. We take preimages of a point~$y\in U$
under~$f$, $f^{-1}(y)$, as $k^j$. Then by summing both sides
of~eq.~(\ref{twentyeight}) over~$j$, we have\footnote{In deriving this
relation, we have assumed that $U$ is within the range of~$f$, i.e., the
inverse image~$f^{-1}(U)$ is not empty. This relation itself, however, is
meaningful even if $U$ is not within the range of~$f$, if one sets $\deg f=0$
for such case. As a consequence, eq.~(\ref{thirtytwo}) holds even if
$f:T^d\to S^d$ is not a surjection, i.e., not an onto-mapping.}
\begin{equation}
   \sum_j
   \int_{U^j}{\cal I}(k;m_0,r)\,\rmd k_1\wedge\cdots\wedge\rmd k_d
   =(\deg f)\int_U\Omega,
\label{thirty}
\end{equation}
where the degree of the mapping~$f$ is given by
\begin{equation}
   \deg f=\sum_{f(k^j)=y}\sgn\left[{\cal I}(k^j;m_0,r)\right].
\label{thirtyone}
\end{equation}
In general, the degree of the mapping $f:T^d\to S^d$ is defined by a sum of the
signature of jacobian of the coordinate transformation between~$T^d$
and~$S^d$ over preimages of a point $y\in S^d$. An important mathematical
fact~\cite{Dubrovin:1984} is that the degree takes the same value for all
(regular) points~$y$ on $S^d$. Thus, by summing eq.~(\ref{thirty}) over all
coordinate patches~$U$ of~$S^d$, we have
\begin{equation}
   I(m_0,r)
   =\int_{T^d}{\cal I}(k;m_0,r)\,\rmd k_1\wedge\cdots\wedge\rmd k_d
   =(\deg f)\int_{S^d}\Omega,
\label{thirtytwo}
\end{equation}
where $\int_{S^d}\Omega$ is given by the volume of the unit sphere~$S^d$:
\begin{equation}
   \int_{S^d}\Omega=\vol(S^d)={2^{d+1}\pi^nn!\over d!}.
\label{thirtythree}
\end{equation}

We may choose any point $y$ on~$S^d$ to evaluate the degree~(\ref{thirtyone}).
We choose $y=(1,0,\cdots,0)$. This requires
\begin{equation}
   x_\mu(k^j)={s_\mu\over\epsilon}=0,\qquad{\rm for}\qquad\mu=1,\ldots,d,
\label{thirtyfour}
\end{equation}
and
\begin{equation}
   x_0(k^j)={m_0+r\sum_\mu(c_\mu-1)\over\left|m_0+r\sum_\mu(c_\mu-1)\right|}
   =1.
\label{thirtyfive}
\end{equation}
Note that eq.~(\ref{thirtyfive}) is equivalent to the
condition~$m_0/r+\sum_\mu(c_\mu-1)>0$. Now eq.~(\ref{thirtyfour}) implies that
$k_\mu^j=0$ or~$\pi$ for each direction~$\mu$. We denote the number of $\pi$'s
appearing in $k^j$ by an integer~$n_\pi\geq0$,
\begin{equation}
   k^j=(\underbrace{\pi,\ldots,\pi}_{n_\pi},0,\ldots,0),
\label{thirtysix}
\end{equation}
irrespective of the position of $\pi$'s. For a given~$n_\pi$, the number of
such $k^j$ is ${d\choose n_\pi}$. The second relation~(\ref{thirtyfive}) on the
other hand requires~$n_\pi<m_0/(2r)$. At those $k^j$, we have
\begin{equation}
   \sgn\left[{\cal I}(k^j;m_0,r)\right]=\prod_\mu c_\mu=(-1)^{n_\pi}.
\label{thirtyseven}
\end{equation}
Thus eq.~(\ref{thirtyone}) gives
\begin{equation}
   \deg f=\sum_{n_\pi=0}^{[m_0/2r]}{d\choose n_\pi}(-1)^{n_\pi}.
\label{thirtyeight}
\end{equation}

Combining eqs.~(\ref{twenty}), (\ref{thirtytwo}), (\ref{thirtythree})
and~(\ref{thirtyeight}), we finally obtain
\begin{equation}
   q(x)={i^n\over(4\pi)^nn!}
   \sum_{n_\pi=0}^{[m_0/2r]}(-1)^{n_\pi}{d\choose n_\pi}\,
   \epsilon_{\mu_1\nu_1\cdots\mu_n\nu_n}
   \tr\left(F_{\mu_1\nu_1}\cdots F_{\mu_n\nu_n}\right)+O(a),
\label{thirtynine}
\end{equation}
where we have used
\begin{equation}
   {-1/2\choose n}={1\over n!}\left(-{1\over2}\right)^n(d-1)!!.
\label{forty}
\end{equation}
In particular, when~$0<m_0/r<2$ with which the (free) overlap-Dirac operator is
free of species doubling~\cite{Neuberger:1998fp}, eq.~(\ref{thirtynine})
reproduces the expected result. Multiplying the volume form
$\rmd^dx=\rmd x_1\wedge\cdots\wedge\rmd x_d$ to $q(x)$, we have
\begin{equation}
   q(x)\,\rmd^dx=\left.\tr\exp\left({i\over2\pi}F\right)\right|_{\rmd^dx}+O(a),
\label{fortyone}
\end{equation}
where the field strength 2-form~$F$ is defined
by~$F=F_{\mu\nu}\,\rmd x_\mu\wedge\rmd x_\nu/2$. This is our main result. When
combined with cohomological arguments, it provides an absolute normalization of
the lattice axial anomaly even with finite lattice
spacings~\cite{Luscher:1999kn,Fujiwara:2000fi} and with finite
volumes~\cite{Igarashi:2002zz}.

\appendix

\section{Axial anomaly with the Wilson-Dirac operator}

With the Wilson-Dirac operator, the chiral symmetry is explicitly broken by the
Wilson term and the axial anomaly is regarded as arising from this explicit
breaking~\cite{Karsten:1980wd}. The axial anomaly thus has a structure of a
fermion one-loop diagram containing the Wilson term and, in our notation, it
reads
\begin{equation}
   q_{\rm w}(x)=-{1\over a^d}\tr
   \left(\gamma_{d+1}{a^2\over2}r\nabla_\mu^*\nabla_\mu
   {1\over A}\right)
   =-{1\over a^d}\tr\left[\gamma_{d+1}{a^2\over2}r\nabla_\mu^*\nabla_\mu
   (A^\dagger A)^{-1}A^\dagger\right],
\label{aone}
\end{equation}
where the parameter~$m_0$ in~$A$ of eq.~(\ref{one}) is replaced by~$am$ with
$m$ being the bare mass of the fermion. An expansion with respect to the
lattice spacing is almost identical to the case of the overlap-Dirac operator
in the text and we have
\begin{equation}
   q_{\rm w}(x)=-{i^n\over(2\pi)^d}\,I_{\rm w}(r)\,
   \epsilon_{\mu_1\nu_1\cdots\mu_n\nu_n}
   \tr\left(F_{\mu_1\nu_1}\cdots F_{\mu_n\nu_n}\right)+O(a),\
\label{atwo}
\end{equation}
where~$I_{\rm w}(r)=\int_{\cal B}\rmd^dk\,{\cal I}_{\rm w}(k;r)$ and
\begin{eqnarray}
   {\cal I}_{\rm w}(k;r)&=&\left(\prod_\mu c_\mu\right)
   \left\{\sum_\nu s_\nu^2+r^2\left[\sum_\nu(c_\nu-1)\right]^2\right\}^{-n-1}
\nonumber\\
   &&\qquad\qquad\times
   \left\{r^2\left[\sum_\rho(c_\rho-1)\right]^2
   +r^2\sum_\rho(c_\rho-1)\sum_\sigma{s_\sigma^2\over c_\sigma}\,\right\},
\label{athree}
\end{eqnarray}
which is independent of the fermion mass. To evaluate this lattice integral, we
again utilize the mapping~(\ref{twentyfour})--(\ref{twentysix}) with $m_0=0$.
The mapping $k_\mu\to x_\mu$ (we do {\it not\/} include $x_0$ here) defines a
mapping~$g:{\cal B}\to B^d$, where $B^d$ is the $d$~dimensional ball,
$B^d=\{x_\mu\in{\mathbb R}^d\bigm|\sum_\mu x_\mu^2\leq1\}$. More precisely, we
have to regard all points of the boundary of the ball, $\sum_\mu x_\mu^2=1$,
are identified, because $k=(0,\ldots,0)$ is mapped to the boundary of the ball.
The crucial relation this time is that the volume form of the ball,
$\rmd x_1\wedge\cdots\wedge\rmd x_d$, coincides with the integrand
of~$I_{\rm w}(r)$ as one can verify. These observations show
\begin{equation}
   I_{\rm w}(r)=(\deg g)\int_{B^d}\rmd x_1\wedge\cdots\wedge\rmd x_d
   =(\deg g)\vol(B^d),
\label{afour}
\end{equation}
where the volume of the $d$~dimensional ball, $\vol(B^d)$, is given by
$\vol(B^d)=\pi^n/n!$ and the degree is
\begin{equation}
   \deg g=\sum_{g(k^j)=y}\sgn\left[{\cal I}_{\rm w}(k^j;r)\right].
\label{afive}
\end{equation}
As the point~$y$ on~$B^d$, we choose the origin $y=(0,\ldots,0)$. All momenta
whose components~$k_\mu$ are either $0$ or~$\pi$, {\it except\/}
$k=(0,\ldots,0)$, are mapped to~$y=(0,\ldots,0)$. Thus the
degree is given by
\begin{equation}
   \deg g=\sum_{n_\pi=1}^d{d\choose n_\pi}(-1)^{n_\pi}=-1.
\label{asix}
\end{equation}
Combining eqs.~(\ref{atwo}), (\ref{afour}) and~(\ref{asix}), we have
$q_{\rm w}(x)\,\rmd^dx=\tr[\exp iF/(2\pi)]|_{\rmd^dx}+O(a)$ for arbitrary
Wilson parameter~$r\neq0$.

\listoftables           % ONLY DRAFT
\listoffigures          % ONLY DRAFT


\begin{thebibliography}{999}

\bibitem{Neuberger:1998fp}
H.~Neuberger,
\emph{Exactly massless quarks on the lattice}, \plb{417}{1998}{141}
[\heplat{9707022}];
%CITATION = HEP-LAT 9707022;%%
%\bibitem{Neuberger:1998wv}
%H.~Neuberger,
\emph{More about exactly massless quarks on the lattice}, \plb{427}{1998}{353}
[\heplat{9801031}].
%%CITATION = HEP-LAT 9801031;%%

\bibitem{Kikukawa:1998pd}
Y.~Kikukawa and A.~Yamada,
\emph{Weak coupling expansion of massless QCD with a Ginsparg-Wilson fermion
and axial $\U(1)$ anomaly}, \plb{448}{1999}{265} [\heplat{9806013}].
%%CITATION = HEP-LAT 9806013;%%

\bibitem{Fujikawa:1998if}
K.~Fujikawa,
\emph{A continuum limit of the chiral jacobian in lattice gauge theory},
\npb{546}{1999}{480} [\hepth{9811235}].
%%CITATION = HEP-TH 9811235;%%

\bibitem{Adams:1998eg}
D.H.~Adams,
\emph{Axial anomaly and topological charge in lattice gauge theory with
overlap-Dirac operator}, \ap{296}{2002}{131} [\heplat{9812003}];
%%CITATION = HEP-LAT 9812003;%%
%\bibitem{Adams:2000rn}
%D.H.~Adams,
\emph{On the continuum limit of fermionic topological charge in lattice gauge
theory}, \jmp{42}{2001}{5522} [\heplat{0009026}].
%%CITATION = HEP-LAT 0009026;%%

\bibitem{Suzuki:1998yz}
H.~Suzuki,
\emph{Simple evaluation of chiral jacobian with the overlap Dirac operator},
\ptp{102}{1999}{141} [\hepth{9812019}].
%%CITATION = HEP-TH 9812019;%%

\bibitem{Chiu:1998qv}
T.-W.~Chiu and T.-H.~Hsieh,
\emph{Perturbation calculation of the axial anomaly of Ginsparg-Wilson
fermion}, \heplat{9901011}.
%%CITATION = HEP-LAT 9901011;%%

\bibitem{Reisz:1999cm}
T.~Reisz and H.J.~Rothe,
\emph{The axial anomaly in lattice QED: a universal point of view},
\plb{455}{1999}{246} [\heplat{9903003}].
%%CITATION = HEP-LAT 9903003;%%

\bibitem{Frewer:2000ee}
M.~Frewer and H.J.~Rothe,
\emph{Universality of the axial anomaly in lattice QCD}, \prd{63}{2001}{054506}
[\heplat{0004005}].
%%CITATION = HEP-LAT 0004005;%%

\bibitem{Luscher:2000un}
M. L\"uscher,
\emph{Weyl fermions on the lattice and the non-abelian gauge anomaly},
\npb{568}{2000}{162} [\heplat{9904009}].
%%CITATION = HEP-LAT 9904009;%%

\bibitem{Alvarez-Gaume:1985ex}
See, for example, L.~Alvarez-Gaum\'e,
\emph{An introduction to anomalies}, HUTP-85/A092
{\it lectures given at Int. School on Mathematical Physics}, Erice, Italy,
Jul 1-14, 1985.

\bibitem{Seiler:1981jf}
E.~Seiler and I.O.~Stamatescu,
\emph{Lattice fermions and theta vacua}, \prd{25}{1982}{2177},
erratum \emph{ibid}.\ {\bf D 26} (1982) 534.
%%CITATION = PHRVA,D25,2177;%%

\bibitem{Aoki:1985hb}
S.~Aoki and I.~Ichinose,
\emph{The Wess-Zumino term in lattice theories}, \npb{272}{1986}{281}.
%%CITATION = NUPHA,B272,281;%%

\bibitem{Aoki:1986tk}
S.~Aoki,
\emph{Lattice fermions and nonabelian anomaly}, \prd{35}{1987}{1435}.
%%CITATION = PHRVA,D35,1435;%%

\bibitem{Haga:1996at}
K.~Haga, H.~Igarashi, K.~Okuyama and H.~Suzuki,
\emph{Remark on Pauli-Villars lagrangian on the lattice},
\prd{55}{1997}{5325} [\hepth{9608025}].
%%CITATION = HEP-TH 9608025;%%

\bibitem{Fujikawa:2000qw}
K.~Fujikawa and M.~Ishibashi,
\emph{Chiral anomaly for a new class of lattice Dirac operators},
\npb{587}{2000}{419} [\heplat{0005003}].
%%CITATION = HEP-LAT 0005003;%%

\bibitem{Coste:1989wf}
A.~Coste and M.~L\"uscher,
\emph{Parity anomaly and fermion boson transmutation in three-dimensional
lattice QED}, \npb{323}{1989}{631}.
%%CITATION = NUPHA,B323,631;%%

\bibitem{Golterman:1992ub}
M.F.L.~Golterman, K.~Jansen and D.B.~Kaplan,
\emph{Chern-Simons currents and chiral fermions on the lattice},
\plb{301}{1993}{219} [\heplat{9209003}].
%%CITATION = HEP-LAT 9209003;%%

\bibitem{Ginsparg:1982bj}
P.H.~Ginsparg and K.G.~Wilson,
\emph{A remnant of chiral symmetry on the lattice}, \prd{25}{1982}{2649}.
%%CITATION = PHRVA,D25,2649;%%

\bibitem{Luscher:1998pq}
M.~L\"uscher,
\emph{Exact chiral symmetry on the lattice and the Ginsparg-Wilson relation},
\plb{428}{1998}{342} [\heplat{9802011}].
%%CITATION = HEP-LAT 9802011;%%

\bibitem{Chandrasekharan:1998wg}
S.~Chandrasekharan,
\emph{Lattice QCD with Ginsparg-Wilson fermions},
\prd{60}{1999}{074503} [\heplat{9805015}].
%%CITATION = HEP-LAT 9805015;%%

\bibitem{Ichinose:1999ke}
I.~Ichinose and K.~Nagao,
\emph{Gauged Gross-Neveu model with overlap fermions},
\emph{Chin.\ J.\ Phys.} {\bf 38}, (2000) 671 [\heplat{9912011}];
%%CITATION = HEP-LAT 9912011;%%
%\bibitem{Ichinose:vj}
%I.~Ichinose and K.~Nagao,
\emph{Gross-Neveu model with overlap fermions},
\mpla{15}{2000}{857};
%%CITATION = MPLAE,A15,857;%%
%\bibitem{Ichinose:1999gs}
%I.~Ichinose and K.~Nagao,
\emph{Lattice QCD with the overlap fermions at strong gauge coupling},
\npb{577}{2000}{279} [\heplat{9910031}];
%%CITATION = HEP-LAT 9910031;%%
%\bibitem{Ichinose:2000cb}
%I.~Ichinose and K.~Nagao,
\emph{Lattice QCD with the overlap fermions at strong gauge coupling 2},
\npb{596}{2001}{231} [\heplat{0008002}].
%%CITATION = HEP-LAT 0008002;%%

\bibitem{Golterman:2000zy}
M.~Golterman and Y.~Shamir,
\emph{Overlap-Dirac fermions with a small hopping parameter},
\jhep{09}{2000}{006} [\heplat{0007021}].
%%CITATION = HEP-LAT 0007021;%%

\bibitem{Giusti:2001xh}
L.~Giusti, G.C.~Rossi, M.~Testa and G.~Veneziano,
\emph{The $\U_A(1)$ problem on the lattice with Ginsparg-Wilson fermions},
\npb{628}{2002}{234} [\heplat{0108009}].
%%CITATION = HEP-LAT 0108009;%%

\bibitem{Hasenfratz:1998ri}
P.~Hasenfratz, V.~Laliena and F.~Niedermayer,
\emph{The index theorem in QCD with a finite cut-off}, \plb{427}{1998}{125}
[\heplat{9801021}].
%%CITATION = HEP-LAT 9801021;%%

\bibitem{Dubrovin:1984}
See, for example,
B.A.~Dubrovin, A.T.~Fomenko and S.P.~Novikov,
\emph{Modern geometry---methods and applications}.
\emph{Part~II. The geometry and topology of manifolds},
Springer-Verlag 1985, \S13 and~\S14.

\bibitem{Luscher:1999kn}
M.~L\"uscher,
\emph{Topology and the axial anomaly in abelian lattice gauge theories},
\npb{538}{1999}{515} [\heplat{9808021}].
%CITATION = HEP-LAT 9808021;%%

\bibitem{Fujiwara:2000fi}
T.~Fujiwara, H.~Suzuki and K.~Wu,
\emph{Noncommutative differential calculus and the axial anomaly in abelian
lattice gauge theories}, \npb{569}{2000}{643} [\heplat{9906015}];
%%CITATION = HEP-LAT 9906015;%%
%\bibitem{Fujiwara:1999fj}
%T.~Fujiwara, H.~Suzuki and K.~Wu,
\emph{Axial anomaly in lattice abelian gauge theory in arbitrary dimensions},
\plb{463}{1999}{63} [\heplat{9906016}].
%%CITATION = HEP-LAT 9906016;%%

\bibitem{Igarashi:2002zz}
H.~Igarashi, K.~Okuyama and H.~Suzuki,
\emph{More about the axial anomaly on the lattice}, \heplat{0206003}.
%%CITATION = HEP-LAT 0206003;%%

\bibitem{Karsten:1980wd}
L.H.~Karsten and J.~Smit,
\emph{Lattice fermions: species doubling, chiral invariance, and the triangle
anomaly}, \npb{183}{1981}{103}.
%%CITATION = NUPHA,B183,103;%%

\end{thebibliography}
\end{document}